\documentclass[twocolumn,aps,floats,floatfix,prl,nofootinbib,superscriptaddress,tightenlines]{revtex4-1}

\usepackage{graphicx}
\usepackage{bm}
\usepackage{epsfig}
\usepackage{pstricks}
\usepackage{amsmath}
\usepackage{hyperref}
\usepackage{color}
\newcommand{\nn}{\nonumber}

\newcommand{\beq}{\begin{equation}}
\newcommand{\eeq}{\end{equation}}
\newcommand{\bea}{\begin{eqnarray}}
\newcommand{\eea}{\end{eqnarray}}
\newcommand{\OMIT}[1]{}

\def\nbarslash{\bar n\hspace{-2mm}\slash}

\newcommand{\nbar}{\bar{n}}

\newcommand{\plusf}[2]{\left [ \frac{#1}{#2} \right ]_{\!+}}
\newcommand{\sqrts}{\sqrt{s}}
\newcommand{\pti}{p_{1t}}
\newcommand{\ptii}{p_{2t}}

\begin{document}

\title{The Rapidity Renormalization Group  }

\def\addCMU{Department of Physics, Carnegie Mellon University, Pittsburgh PA  15213, USA}
\author{Jui-yu Chiu}
\author{Ambar Jain}
\author{Duff Neill}
\author{Ira Z. Rothstein}
\affiliation{\addCMU}

\begin{abstract}

We introduce a systematic approach  for the resummation of  perturbative series 
which involve large logarithms  not only due to large invariant mass ratios but
large rapidities as well. Series of this form can appear in a variety of gauge theory observables.
The formalism is utilized to calculate the jet broadening event shape in a systematic
fashion to next to leading logarithmic order. An operator definition
of the factorized cross section  as well as   a closed form of the next-to leading log
cross section are presented.  The result agrees with the data to within
errors.
\end{abstract}

\maketitle 
Observables in  weakly coupled gauge theories
often  necessitate perturbative resummations to be
under calculational control. This need arises when one performs measurements that are sensitive to infrared scales. By probing  distances  long compared
to the hard scattering scale one introduces large logarithms (logs) that lead to the breakdown
of fixed order perturbative series. Resumming the large logs has  become  standard
in QCD \cite{sterman} and  can be accomplished by factorizing the cross section into momentum regions. Factorization makes 
clear the distinction between logs of various ratios that may be involved in the observable and
resummation follows via standard renormalization group techniques.

An elegant formalism for  factorization is SCET (Soft-Collinear Effective Theory) \cite{SCETI}, which is an
effective field theory designed to reproduce the infra-red physics in high energy processes.
The formalism not only streamlines factorization proofs \cite{SCETII}, but also allows
one to systematically include power corrections. The results of this paper will all be couched
in terms of this framework.

A generic factorized cross section takes on the form:
\beq
\sigma= H\otimes [\Pi_{i} J_i ]\otimes S.
\eeq
The hard function $H$  is responsible for reproducing the short distance physics with wavelengths of
order $1/Q$, where $Q$ is the scale involved in the hard scattering.  $J_i$  and $S$  are the so-called jet  and soft functions containing modes which are highly energetic (collinear)  and soft, respectively. 
Soft modes have small rapidities ($k_+/k_-\sim 1$) while the rapidities of collinear modes are parametrically larger
($k_\pm/k_\mp \gg 1$), where $k_{\pm}$ are the light cone momenta.
The tensor product implies the existence of one or more convolution in momentum space.
In canonical situations, the resummation of large logs is accomplished by evolving, via the renormalization group (RG), each factorized component to its natural scale.   The natural scales are set by the arguments of the logs. $H,J$ and $S$ may contain, for example, logs of  $Q/\mu, m_J/\mu, m_S/\mu$, respectively, where $m_J$ and $m_S$ are quantities  which probe the  invariant masses of the modes composing $J$ and $S$.

While there is a large disparity in rapidity between  the modes which compose $S$ and $J$, the typical invariant mass of the modes need not  be distinct.  When  soft fields have invariant mass parametrically smaller than the collinear modes (in this case the soft mode are called ``ultra-soft" ) significant simplifications arise. Whether or not there is a distinction in invariant
masses, one must always ensure that there is no double counting of modes. That is, loop integrals
within a prescribed function ($J$ or $S$) should only account for the relevant mode. In principle this
could be accomplished using a cut-off, but this would lead to multiple technical difficulties
not the least of which is the need for gauge non-invariant counter-terms.  
Within the effective field theory formalism using dimensional regularization, this double counting is avoided by the so-called
zero-bin procedure \cite{zerobin}. In this methodology one subtracts  from each loop integral
its value when the integrand is asymptotically expanded around the extraneous region. This procedure not only formally avoids
the double counting, but also ensures that all integrals in the theory are well defined.
Moreover, the zero bin subtraction, or some equivalent subtraction method,  is necessary to preserve factorization \cite{Chiu:2009yx}.
This potential breakdown of factorization occurs as a consequence of the need to regulate 
``rapidity divergences" (light-cone singularities). These divergences arise schematically from integrals of the
form 
\beq
I_R= \int \frac{d k_+}{k_+},
\eeq
which are not regulated in dimensional regularization.  There are multiple ways of regulating
this divergence. A simple way is to introduce a new dimensionful parameter $\Delta$ via 
the replacement $k_+ \rightarrow k_++\Delta$ in the denominator. Note that regulating
these divergences will break boost invariance along light-cone direction.  For any physical observable the final result must
be  boost invariant and  independent of $\Delta$,  this is automatic once zero-bin subtraction is performed and all sectors  added.
For the case of $m_J \gg m_s$, $\Delta$ dependence cancels in each sector after zero bin subtraction, and the boost symmetry is restored in each sector. This will, however, not be the case when $m_J \sim m_S$, since the soft and collinear modes mix under boosts.

When $m_J \sim m_S$  the jet \emph{axis}, defined as the direction of the net momentum of the
jet, recoils against the soft emission.   In the light-cone co-ordinates, collinear modes scale like $(n \cdot p,\nbar \cdot p , p_\perp) \sim Q(1,\lambda^2,\lambda)$ while soft modes scale as $Q(\lambda,\lambda,\lambda)$, where  $\lambda\ll 1$ and $n^\mu$ is the light-cone direction chosen to perform the factorization.
The jet axis is no longer aligned with $n^\mu$,
and  one should not expect the jet function to be invariant
under boosts along $n^\mu$. 
However, the sum of all sectors will still be invariant. If we were to reanalyze the rapidity divergences discussed previously we would find that the $\Delta$ regulator will cancel
after summing over contributions from all sectors (with proper zero-bin subtractions) \cite{Chiu:2009yx}.
Boost non-invariance in the jet function appears in the form  $\log[Q/\Delta]$, a ``rapidity log".
 Resumming these  logs using a $\Delta$ regulator, is technically cumbersome  much like 
 resummations with a cut-off regulator. Here we introduce a regulator more in the spirit of dimensional regularization that does not introduce new dimensionful scales in the integrals, and maintains manifest power counting
 in the effective theory.

\begin{figure}[t!]
{\includegraphics[width=0.4\textwidth]{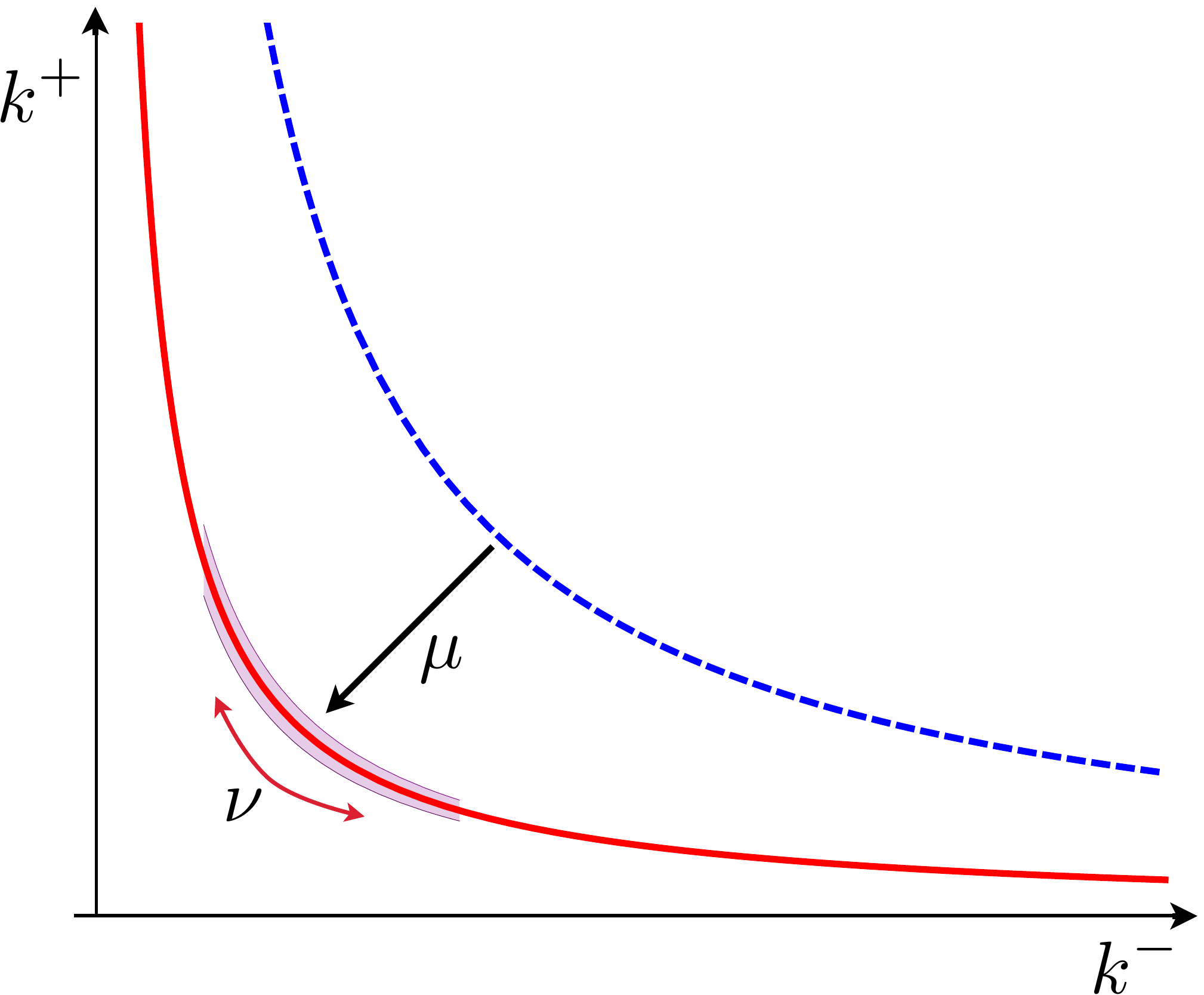}}\vskip-0.3cm
\caption[1]{ \label{fig1}Rapidity renormalization group flow  along the on-shell hyperbola versus the standard flow.
}
\end{figure}

Given the existence of the rapidity logs in addition to the canonical logs, 
 $S$ and $J$ may not have one definitive scale associated with them.
To resum both sets of logs we will introduce another arbitrary scale $\nu$, along the
lines of $\mu$ in dimensional regularization. 
We expect that in order to properly resum all the large logs we will need to
run the jet in $\nu$ down to the smaller rapidity scale of the soft function.
In a Wilsonian sense we have  two distinct cut-offs with which to thin degrees of
freedom. There will be one flow in invariant mass and one in rapidity as shown in Fig.~\ref{fig1}. This is an inherently Minkowskian procedure.

To illustrate this rapidity renormalization group we will consider the specific example of the event shape called jet broadening.
 Event shapes have  played an important role in precision measurements
of the strong coupling $\alpha_s$\cite{OPAL}.  A generalized event shape for event $e^-e^+\rightarrow X$ at center of mass energy $\sqrts$,  can be defined \cite{berger} in terms
of a parameter $a$ via
\beq
\label{ang}
e(a)= 
\sum_{i\in X}
 \frac{\vert \vec p_{i\perp} \vert }{\sqrts}e^{-\vert \eta_i\vert(1-a)}
\eeq
where $p_{i\perp}$ is the transverse momentum with respect to the thrust axis of the event, and $\eta_i$ is the rapidity of the
i'th particle. The thrust axis $\hat t$ is defined via ${\rm max}_{\hat t}  \sum_{i \in X}\vert  p_i \cdot \hat t \vert/\sqrts$.
Two particularly interesting event shapes are the limits $a=0,1$ corresponding to ``thrust'' and ``jet broadening''
respectively. The limit $e\ll 1$ isolates events composed of back to back jets.  In the case of
thrust these jets are composed of collinear radiation, and the recoil due to soft (ultra-soft in this case) radiation does not affect the jet axis. While for jet broadening all radiation with parametrically similar transverse momentum can contribute, so that soft radiation of the form $Q(\lambda,\lambda,\lambda)$  recoils the jet off the thrust axis. In both of these cases fixed order perturbation theory will fail
when $e$ is small. However, as long as
$eQ\gg \Lambda_{QCD}$, we expect non-perturbative effects to be suppressed, though large logs of $e$ need to be resummed.

 The pioneering work on jet broadening resummations \cite{catani1}
utilized the coherent branching  formalism \cite{catani2}. It was later stated \cite{Salam} that
the results in \cite{catani1} neglected terms due to recoil of soft gluons.
In this letter we will provide a factorization theorem for jet broadening, whose proof will follow in a subsequent publication \cite{future}.
 The factorization proofs for  angularity observables (\ref{ang}) in \cite{bauer}
are known to fail as $a$ approaches $1$, since there are  growing power corrections in this limit where one approaches jet broadening. The reason for the apparent breakdown of factorization is the fact that in this limit the soft radiation has the same invariant mass as collinear radiation and
one must change the power counting  accordingly to factorize in a consistent fashion \cite{future}.

Hereforth, we set $a=1$ and $e(1)\equiv e$. In \cite{future} we prove that the cross section for jet broadening takes the following factorized form
\begin{align}
\label{fact}
&\frac{d\sigma}{de}= \sigma_0 H(s) \int de_{n}de_{\nbar} de_{s} \delta(e-e_{n}-e_{\nbar}-e_{s}) \\
& \int d{p}_{1t}d{p}_{2t}
J_{n}(Q_+,e_n,p_{1t})J_{\nbar}(Q_-,e_{\nbar},p_{2t}) {\cal S}(e_s,p_{1t},p_{2t}) \nn
\end{align}
where in covariant gauges 
\begin{align}
& J_n\!\!= \Omega_{\bar d}\!\!\int \!\!\frac{dx_+}{2N_c} e^{\frac{iQ^-x^+}{2}}  
\langle 0\!\!\mid \bar \chi_n(x_+)\frac{\nbarslash}{2}\delta(\hat e -e_n) \delta(\hat P_\perp+\vec p_{1\perp}) \chi_n (0)\!\!\mid \!0\rangle, \nn \\
&{\cal S}= p_{1t}^{1-2 \epsilon} p_{2t}^{1-2 \epsilon} \Omega_{\bar d}\int \frac{d\Omega_{12}}{N_c}  \times \\
& \!\!\langle 0\!\!\mid  S_n^\dagger S_{\nbar} \delta(\hat e -e_s) \delta^{\bar d}(\hat P_{n\perp}- \vec p_{1\perp}) \delta^{\bar d}(\hat P_{\nbar\perp}-\vec p_{2\perp})S_{\nbar}^\dagger S_n\!\!\mid \!0\rangle, \nn
\end{align}
and $\sigma_0$ is the Born cross section. $H$ is the hard matching coefficient which incorporates all the short distance contributions, and is fixed
by matching the QCD currents onto the SCET currents.  Here, $\bar d= 2-2\epsilon$, 
$\chi_{(n,\nbar)}$ are gauge invariant \ SCET fields which include collinear Wilson lines $W_{n,\bar n}$ and the $S_{n,\nbar}$ are soft wilson lines. $\hat P_{n\perp}$ and $\hat P_{\nbar\perp}$ are hemisphere-transverse momentum operators, $\Omega_{\bar d}$ refers to area of a $\bar d$-dimensional unit sphere, $\Omega_{12}$ the relative angle between $\vec p_{1\perp}$ and $\vec p_{2\perp}$, and $p_{it} = \vert \vec p_{i\perp}\vert$. Finally, $Q^\pm$ are the large light cone momenta of the jets with constraint $Q^+ Q^- = s$, the center of mass energy.

As long as $\sqrts e\gg \Lambda_{QCD}$ all of these matrix elements are calculable in perturbation theory.
The bare matrix elements possess both rapidity and UV divergences. Thus standard dimensional regularization is insufficient to regulate all the integrals.  Beyond tree level  one is immediately met with the aforementioned rapidity divergences.
To regulate these integrals we introduce a regulator into the momentum space SCET Wilson lines as follows
\begin{align}
\nonumber&\!\!\! W_n = \Bigg [\sum_{\rm perm} \exp \left ( \frac{-g}{\bar n\!\cdot\! \hat P} \, \left [w  \frac{ \vert \bar n \!\cdot\! \hat P\vert^{-\eta}}{\nu^{-\eta}} \,\bar n\! \cdot \! A_{n,q}(0) \right ] \right )\Bigg ] \, ,\\
& \!\!\! S_n = \Bigg [\sum_{\rm perm} \exp \left (\frac{ -g}{n\!\cdot\! \hat P} \, \left [w \frac{ \vert  2 \hat P^0 \vert^{-\eta/2}}{\nu^{-\eta/2}} \, n\! \cdot \! A_{s,q}(0) \right ] \right )\Bigg ]\, ,
\end{align}
where $\nu$ is an arbitrary scale independent of the usual scale $\mu$ introduced
in dimensional regularization. $w$ is a bookkeeping parameter which will be set to one
at the end of the calculation. $\hat P^\mu$ is the momentum operator and we have essentially regulated
the energy of the emitted gluons in each Wilson line,  since $2P_0 \to \nbar \cdot P$ in the collinear limit. 
 Notice the factor of $\eta/2$ in the soft function.
This choice is not a matter of convention. Physically, the factor  arises as a consequence  of the fact that soft must be cut-off at both 
positive as well as negative rapidity \cite{future}. 
 The Wilson line regularization breaks manifest boost symmetry, 
 which is restored 
once all of the sectors are combined. The rapidity divergences  for the jet and soft functions will introduce a new
set of anomalous dimensions  $(\gamma^\nu_J,\gamma^\nu_S) $which are defined via variation of $\nu$.
Given that the hard function has no such anomalous dimensions, we must have the relation
\beq
2 \gamma^\nu_J +\gamma^\nu_S=0,
\eeq
or equivalently it must be true that the total $\eta$ dependence must vanish.
Indeed it is not hard to show \cite{future}, that the sum of the $\eta$ divergences cancels
as a consequence of   eikonalization. This cancellation also implies that the individual factors
$J$ and $S$ are multiplicatively renormalizable.

The tree level  jet function is given by $\delta(e_n- p_{1t}/\sqrts)$ and soft function by $\delta(e_s)\delta(\pti)\delta(\ptii)$.
To determine the relevant scales in the logs, we must convolve the renormalized one-loop jet function 
with the tree level soft function as dictated in (\ref{fact}).
This is most easily seen for the integrated cross-section $\Sigma = \int_0^{e_0} de\, (d\sigma/de)$. At order $\alpha_s$, the result of the convolution leads to the following singular contributions from a jet 
\begin{multline}
\frac{1}{\sigma_0}\Sigma_{\rm jet}= \frac{-\alpha_sC_F}{2\pi} \ln(\frac{\sqrts e_0}{2\mu}) \Big[3
+4\ln(\frac{\nu}{Q^{\pm}})\Big] \, .
\end{multline}
It thus becomes clear that the jet function depends on two different kinematical scales ($\sqrts e_0$ and $Q^{\pm}$).
In addition, we see that dependence on rapidity manifests in the form of the rapidity log, 
 $\ln(Q^{\pm}/\nu)$.
While the soft function singular contributions are \begin{align}
\!\!\!\!\!  \frac{1}{\sigma_0}\Sigma_{\rm soft}=\frac{\alpha_sC_F}{\pi}\Big[ -2 \ln^2(\frac{\sqrts e_0}{2\mu})+4\ln\frac{\nu}{\mu}\ln(\frac{\sqrts e_0}{2\mu})\Big],
\end{align}
and has rapidity logs set at the low scale $\sqrts e_0$.

\begin{figure}[t]
\includegraphics[width=0.3\textwidth]{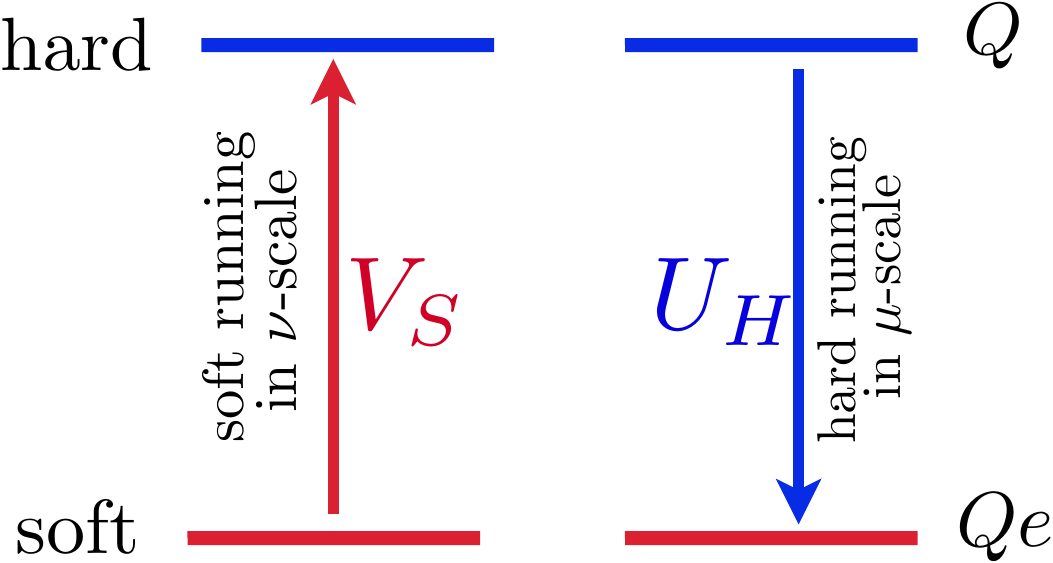}
\caption{Running Strategy.}
\label{fig2}
\end{figure}

The utility of $\nu$ is clear
as we may choose $\nu \sim Q^{\pm}\sim \sqrts$ and $\mu \sim \sqrts e$ to minimize the logs in the jet function. Then to minimize the logs in the
soft function we run $\nu$ from the scale $\sqrts e$ up to  $\sqrts$. Furthermore we will need to run the hard matching coefficient down
to the scale $\sqrts e$ in the dimensional regularization parameter $\mu$.
This scenario is shown schematically in Fig.~\ref{fig2}.

Here we will perform the running at next to leading log (NLL) , which sums all terms of order one, where
we take the scaling $\alpha_s \ln(e) \sim 1$. The NNLL analysis will be performed
in \cite{future}.
The hard function RGE is well known to NNLL (see \cite{hard}).  
The running is most simply performed in Laplace transform space, with ($b,b^\prime)$ conjugate to ($\pti,\ptii)$ respectively.
We find
\begin{align}
\gamma_\nu^S(b,b^\prime)= -2\Gamma_{c}[ \left( \log (b \mu e^{\gamma_E})+\log (b^\prime \mu e^{\gamma_E}) \right ] \, ,
\end{align}
where $\Gamma_c$ is the usual cusp anomalous dimension for Wilson lines. Similar equations can be written for the two jet functions in terms of their corresponding anomalous dimensions. 


In our strategy of resumming logs  for the soft function we only need two-loop cusp. Solution of the $\nu$-RGE for the soft function is
\begin{align}
{\cal S}(\mu,\nu) = V_s(\mu,\nu/\nu_0) \otimes {\cal S}(\mu,\nu_0) \, 
\end{align}
where $\otimes$ represents convolution in kinematical arguments which are dropped for brevity. Here,
\begin{align}
&V_s\big(\pti,\ptii; \omega_s(\mu,\nu/\nu_0)\big) = \frac{e^{-2\gamma_E \omega_s}}{\Gamma^2(1+\omega_s)} \\
& \Bigg( \frac{\omega_s}{\mu} \plusf{1}{(\frac{\pti}{\mu})^{1-\omega_s}} \!\!\!+ \delta(\pti) \Bigg) \Bigg( \frac{\omega_s}{\mu} \plusf{1}{(\frac{\ptii}{\mu})^{1-\omega_s}} \!\!\!+ \delta(\ptii) \Bigg) \nonumber \, ,
\end{align}
with
$
\omega_s(\mu,\nu/\nu_0) = 2 \Gamma_{\rm c}[\alpha_s(\mu)]  \,  \log \frac{\nu}{\nu_0} \, .
$
To minimize logs in the hard function we need to evolve the hard function using
\begin{align}
H(s;\mu) = H(s;\mu_0) \, U_H(s;\mu0,\mu) \,
\end{align}
where up to NLL $U_H$ can be found in \cite{hard}, and $H(s,\mu_0)=1$ to the order we are working.

The results we have presented so far are for the angularity at $a=1$, which is related to the total jet broadening $B_T$ via $e = 2 B_T$. We will present cross-sections for total jet broadening here and compare with the data. For NLO singular cross-section we get
\begin{align}
\frac{d\sigma}{dB_T} = \sigma_0 \frac{\alpha_s(\mu)C_F}{\pi \, B_T} \left ( - 3 - 4 \log B_T \right ) \, 
\end{align}
where $\sigma_0$ is the Born cross-section. This result is in agreement with ref.~\cite{catani1}. For the re-summed cross-section up to NLL order we have
\begin{align}
\frac{d\sigma}{dB_T} = \frac{\sigma_0}{B_T} \frac{U_H(Q^2, \mu_Q, \mu)}{\Gamma(2\omega_s)e^{2\gamma_E \omega_s}}\left(  \frac{Q B_T}{\mu} \right)^{2\omega_s} \, .
\end{align}

\begin{figure}[t]
\includegraphics[width=0.49\textwidth]{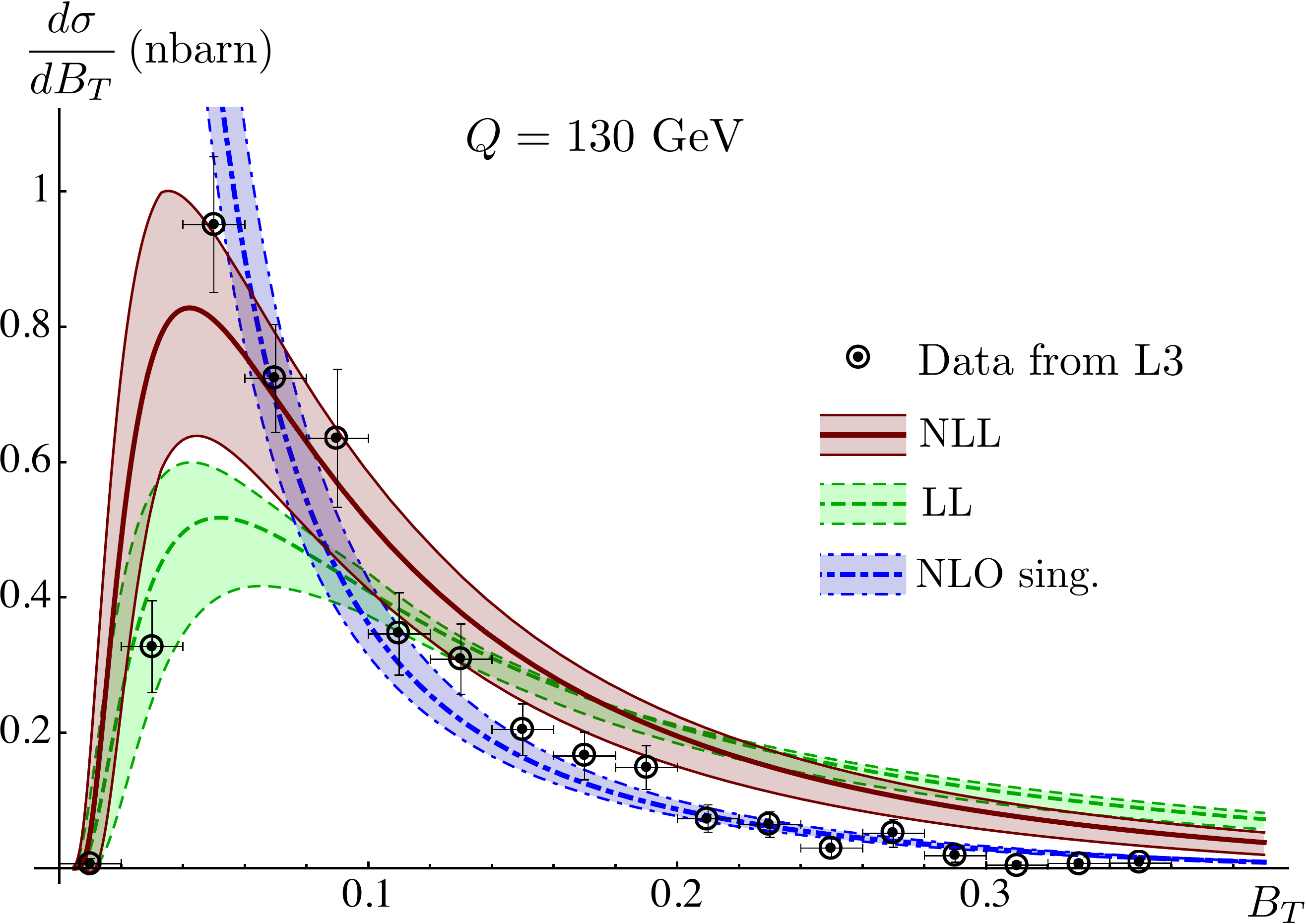}
\caption{Total Jet Broadening at 130 GeV.}
\label{fig3}
\end{figure}
In Fig.~\ref{fig3} we have plotted the theory cross section and the data \cite{data}.
We see that given the large error bars the agreement with the data is reasonable.
However, the NNLL calculation will reduce the theory errors considerably.
We have not included the theory errors due to power corrections.
In the small $B_T$ region these are non-perturbative and scale as $\Lambda_{QCD}/(B_TQ)$ and can be expected to be of order $20$-$30$\%. 
In the tail region there are corrections of order $B_T$ relative to the singular contributions. The disagreement at intermediate values of $B_T$, where fixed order calculations suffice, is expected, since logs will not dominate in this region and 
NLL results leave off order one contributions.
This region will be correctly reproduced in the NNLL calculation. Therefore, by systematically improving this result by including higher order corrections in $\alpha_s$, power corrections and non-perturbative correction, this result can be used for precision $\alpha_s$ determination. Such an analysis using thrust was done in \cite{hard}.

Finally, we wish to point out that the rapidity renormalization group can be utilized in multiple
other settings where rapidity divergences arise. Generically, this will occur whenever kinematically
soft radiation has invariant mass of the same order as the collinear radiation, as in 
cases where one measures the $p_T$ of the final state. Such observables
will be discussed in more detail in \cite{future}. Furthermore, it would be interesting
to utilize our rapidity renormalization group in the context of exclusive processes 
where it has been shown rapidity factorization sheds light on end point singularities
in integrals over light-cone distribution functions \cite{zerobin}.

We would like to thank Tom Ferguson for useful discussions regarding the data. We also thank I.W. Stewart for discussions. This work is supported  by DOE Grants DOE-ER-40682-143 and DEAC02-
6CH03000. D.N. is supported by LHC-TI Graduate Fellowship NSF grant number PHY-0705682.

\end{document}